\documentclass{aa}

\usepackage{graphics}

\def\gtaprx {\lower .1ex\hbox{\rlap{\raise .6ex\hbox{\hskip .3ex
             {\ifmmode{\scriptscriptstyle >}\else
                {$\scriptscriptstyle >$}\fi}}}
                \kern -.4ex{\ifmmode{\scriptscriptstyle \sim}\else
                {$\scriptscriptstyle\sim$}\fi}}}

\begin{document}

\thesaurus{03(11.01.02; 11.17.3; 13.09.2)}
\title{ISOPHOT\thanks{ISO is an 
 ESA project with instruments funded by ESA member states (especially
 the PI countries: France, Germany, the Netherlands and the United
 Kingdom) and with participation of ISAS and NASA} observations of 3CR
 quasars and radio galaxies}

\author{Ilse M. van Bemmel\inst{1,2} \and Peter D. Barthel\inst{2} \and
Thijs de Graauw\inst{3}}

\institute{European Southern Observatory, Karl-Schwarzschildstr. 2,
           D--85748 Garching bei M\"unchen
        \and
        Kapteyn Astronomical Institute, P.O. Box 800, NL--9700~AV
           Groningen, The Netherlands
        \and
        Dutch Space Research Organization, P.O. Box 800, NL--9700~AV
           Groningen, The Netherlands}

\offprints{Ilse van Bemmel (ivbemmel@eso.org)}

\date{Received date; accepted date}

\authorrunning{van Bemmel et al.}
\titlerunning{}

\maketitle

\begin{abstract}

In order to check for consistency with the radio-loud AGN unification
scheme, ISOPHOT data obtained for two small sets of intermediate
redshift steep-spectrum 3CR radio galaxies and quasars are being
examined.  Supplementary submillimeter and centimeter radio data for the
quasars are also taken into account, in order to assess the magnitude of
any beamed nonthermal radiation.  
The fact that we find broad-lined objects to be somewhat more luminous in
their far-infrared output than narrow-lined objects, hints at a
contradiction to the unification scheme. However,
as the sample objects are not particularly well
matched, the sample size is small, and the FIR radiation may still be
partly anisotropic,
this evidence is, at the moment, weak.


\end{abstract}

\section{Introduction}

Several arguments from radio astronomy suggest that all radio-loud
quasars are oriented towards us (Barthel 1989) and that these quasars
should be identified with favourably oriented luminous radio galaxies
such as Cygnus~A.  The unified theory for radio-loud active galaxies
(e.g.  Urry \& Padovani 1995) indeed states that different types of
powerful (Fanaroff \& Riley 1974 class~II) extragalactic radio sources are
actually the same objects, but seen at different orientation angles. 
This orientation dependence is caused by an opaque dusty torus that
surrounds the central engine and thus blocks certain types of radiation
in certain directions.  This dust torus must, however, be transparant to
radio, submillimeter and hard X-ray radiation.  Recent work has indeed
revealed the X-ray nucleus in Cygnus~A (Ueno et al. 1994), as well as
direct (Tadhunter et al. 1999) and scattered (Ogle et al. 1997)
optical and near-infrared signatures from its hidden quasar. 

The opaque torus -- postulated earlier for Seyfert galaxies (e.g. 
Antonucci 1993) -- is believed to absorb most of the hard non-thermal
radiation emanating from the central engine, and must reradiate the
energy at infrared wavelengths. In the past years several models were
developed for this process with different approaches, but with
comparable results. With increasing optical depth, the torus' far
infrared radiation becomes more dependent on viewing angle due to the
aspect geometry. For moderately thick tori however, this dependence
disappears for wavelengths in excess of $\sim60\,\mu$m (Granato \&
Danese 1994), where the dust becomes optically thin. The models of
Pier \& Krolik (1992) cover a larger range of optical depths and here
the anisotropy can be sustained up to $\sim80\,\mu$m. 

%
%
\begin{table*}[!ht]
\label{chartab}
\begin{center}
\leavevmode
\footnotesize
\begin{tabular}[h]{lrrrr|lrrrr}
\hline \\[-8pt]
Quasar & $z$ & $P_{178}$ (W/Hz) & size (\arcsec) & Obs. date & Radio gal. &
 $z$ & $P_{178}$ (W/Hz) & size (\arcsec) & Obs. date \\
\hline \\[-8pt]
\object{3C\,334}   & 0.555 & 27.88 & 58 & 17 Aug '97 & 
\object{3C\,19}    & 0.482 & 27.75 & 10 & 22 Jul '97 \\
\object{3C\,351}   & 0.362 & 27.57 & 75 & 18 Jul '97 & 
\object{3C\,42}	   & 0.395 & 27.51 & 28 & 12 Jul '97 \\
\object{3C\,323.1} & 0.264 & 27.13 & 69 &  8 Aug '97 & 
\object{3C\,460}   & 0.268 & 27.08 & 8 &  2 Dec '97 \\ 
\object{3C\,277.1} & 0.321 & 27.24 & 1.7 &  6 Jul '97 & 
\object{3C\,67}	   & 0.310 & 27.28 & 2.5 & 19 Jul '97 \\
\hline \\[-8pt]
\end{tabular}
\end{center}
\caption{Pair sample characteristics and dates of the ISOPHOT
observations. References to radio images: Bridle et al. (1994) --
3C\,334, 3C\,351; Sanghera et al. (1995) -- 3C\,277.1, 3C\,67; Bogers
et al. (1994) -- 3C\,323.1; Jenkins et al. (1977) -- 3C\,19; Fernini
et al. (1997) -- 3C\,42; Laing (1980) -- 3C\,460. Radio source power 
values were computed using H$_0$=75 km~s$^{-1}$~Mpc$^{-1}$ and q$_0$=0.5.}
\end{table*}
\normalsize

Combining the unification theory with the hypothesized properties of the
dust torus, it follows that the long wavelength far-infrared flux of
radio-loud active galactic nuclei (hereafter AGN) should not depend on
their identification as quasar or radio galaxy. Blazars form an
exception, since for these objects most, if not all, far-infrared
radiation is beamed non-thermal radiation. Hence, far-infrared
photometry would classify as a good consistency check for unification
models. IRAS observations of powerful double-lobed 3CR quasars and
radio galaxies at intermediate redshift revealed that the former class
is somewhat brighter at 60\,$\mu$m than the latter (Heckman et al.,
1994, Hes et al. 1995). However, noting that the restframe wavelength
for these sources lies around 40\,$\mu$m, the torus models can still
account for this infrared excess as it is likely that the torus' optical
depth still exceeds unity at $\sim40\,\mu$m. In addition, a non-thermal
component is expected to play a role. As mentioned above, beamed
non-thermal radiation dominates the overall spectral energy
distributions of blazars (e.g., Impey \& Neugebauer 1988). In the
framework of unification part of this beamed non-thermal emission could
be observable in quasars and thus boost the far-infrared output. In
radio galaxies such a component is not visible, due to their
perpendicular orientation to the line of sight. Although this has been
proposed as a possible solution by Hoekstra et al. (1997), the actual
amount of non-thermal far-infrared emission in double-lobed quasars
appeared rather small (van Bemmel et al. 1998). 

Detailed measurements of the far-infrared--submm spectral energy
distributions of powerful AGN are still sparse. ISO observations by
Rodr\'{\i}guez~Espinosa et al. (1996) and most recently Haas et al. 
(1998) show that at least three emission components can be isolated. 
In addition to the beamed nonthermal component, thermal emission from
AGN related warm ($\sim$100--600K) and starburst related cool
($\sim$20--50K) dust is measured. The present paper is primarily concerned
with the former but we note that also the latter can be strong in
powerful radio sources. For instance, quasar \object{3C\,48} -- known
to be hosted by a gas rich merger (Stockton \& Ridgway 1991, Wink et
al. 1997) -- displays an unusually luminous cool dust component (Hes et
al. 1995, Haas et al. 1998). 

To uncover the cause of the far-infrared excess in quasars with respect
to their radio-galaxy counterparts, and to assess the nature of the
emission, observations longward of 60\,$\mu$m are needed.  Current torus
models predict that beyond 80\,$\mu$m the thermal emission will be
isotropic. Thus, quasars and radio galaxies of comparable AGN strength
should emit comparable amounts of long wavelength far-infrared emission. 
We report here on an observing program with ISOPHOT (Lemke et al. 1996)
to obtain photometry at three far-infrared wavelengths of 3CR quasars
and radio-galaxies.  In order to quantify any non-thermal
contamination, we also observed our sample quasars in the submm with the
JCMT and at cm wavelengths with the NRAO Very Large Array (VLA). 

\section{Sample selection and observations}

The prerequisite for a fair comparison of the reprocessed AGN radiation
is that the sample objects have comparable AGN strength.  We originally
proposed to observe pairs of steep radio spectrum, intermediate redshift
3CR quasars and radio galaxies, matched in radio (lobe) power and
redshift.  The orientation independent, long wavelength radio lobe power
is considered to be a reasonably good measure of the central engine
power (Willott et al. 1999, Rawlings \& Saunders 1991).  Due to
relativistically beamed radiation, and consistent with the unification
model, quasar radio cores are generally somewhat brighter than radio
galaxy cores, at short (cm) wavelengths. 

The original sample also included Compact Steep Spectrum (CSS) radio
sources, having subgalactic dimensions. These objects are believed to
be young objects, still in the phase of radio lobe expansion (e.g.
De Vries et al. 1998). Inclusion of these objects is targeted at
obtaining information on any possible evolution in infrared emission of
radio sources with source age. On the basis of radio morphological
parameters, Fanti et al. (1990) finds powerful CSS quasars and radio
galaxies to be consistent with orientation unification. 

Following acceptance of this project within the European Quasar Core
Programme, ISOPHOT observations were scheduled.  Only after a long
period of observing mode and calibration strategy testing by the
instrument team, we finally observed a subset of our original samples,
at fewer wavelengths than planned, during the period July to December
1997.  We eventually decided to make raster-mode mini-maps with the C100
and C200 detectors, using mode P22.  The rasters have a size of
$3\times3$ pixels for C100 and $4\times2$ pixels for C200 in Y$\times$Z
direction.  This raster technique produced final maps with a size of
$4\arcmin\times4\arcmin$ and $7.5\arcmin\times4.5\arcmin$,
respectively.  The filters
used were the 60\,$\mu$m, 90\,$\mu$m (C100) and 160\,$\mu$m (C200). 
From its original conception in 1993 to its completion in 1997, this
project was, unfortunately, considerably reduced in scope by the
instrument limitations; only four quasar -- radio galaxy pairs were
observed by ISO before the end of its lifetime.

In order to detect possible non-thermal contamination of the
far-infrared emission of the quasars, their core flux densities were
measured with the VLA and JCMT.  These observations were done close in
time to the ISO measurements, in order to minimize the effects of core
variability.  VLA A-array data were obtained at four wavelengths: 6, 2,
1.2 and 0.7\,cm, on April 17, 1998.  JCMT SCUBA data were obtained at
three wavelenghts: 2\,mm, 850\,$\mu$m and 450\,$\mu$m on May 28, 1997. 
Since the effect of non-thermal emission in weak-core radio galaxies is
negligible (Hoekstra et al. 1997), these were not observed with the VLA
and SCUBA.  SCUBA data could not be obtained for quasar 3C\,351, due to
its high declination. 

Table~1 specifies the characteristics of the objects, together with the
dates of the ISO observations.  Three of the pairs consist of Fanaroff
\& Riley class~II objects (Fanaroff \& Riley 1974) with large,
double-lobed, edge-brightened radio morphologies.  It should be noted
that in these three pairs the quasars are systematically larger in
projected size than the paired radio galaxies.  The fourth pair is
comprised of two subgalactic size radio sources of the CSS class.  We
will return to the pair properties in the discussion section.

\section{Data reduction}

\subsection{ISOPHOT data reduction}

All ISO data have been processed with OLP version 8.4 and raw data have
been reduced using the Phot Interactive Analysis tool (PIA) version 8.0. 
The PIA dataflow is divided into four levels, each having an increased
amount of reduction performed.  The Edited Raw Data (ERD) level contains
the raw data in Volts, read directly from the detector and plotted
against time.  At this level the ramps are linearized and a first
deglitching is performed to remove cosmic hits.  The two-threshold method
is used, with a sigma of 3.0 for flagging and a sigma of 0.5 for
reacceptance.  The data is then processed into Signal per Ramp Data
(SRD) level by fitting the ramps with a first order polynomial.  To
check the data for remaining tails from cosmic hits, we use ramp
subdivision.  The ramps are divided in parts containing at least 8 or 16
read-outs and each part is fitted with a separate first order
polynomial.  However, since this dramtically increases the noise, this
is only done as a test of the two-treshold deglitching accuracy and the
final processing is done without subdivision.  The results do not
differ, meaning the deglitching at ERD level is of good quality. 

%
%
\begin{figure*}
  \resizebox{18cm}{!}{\includegraphics{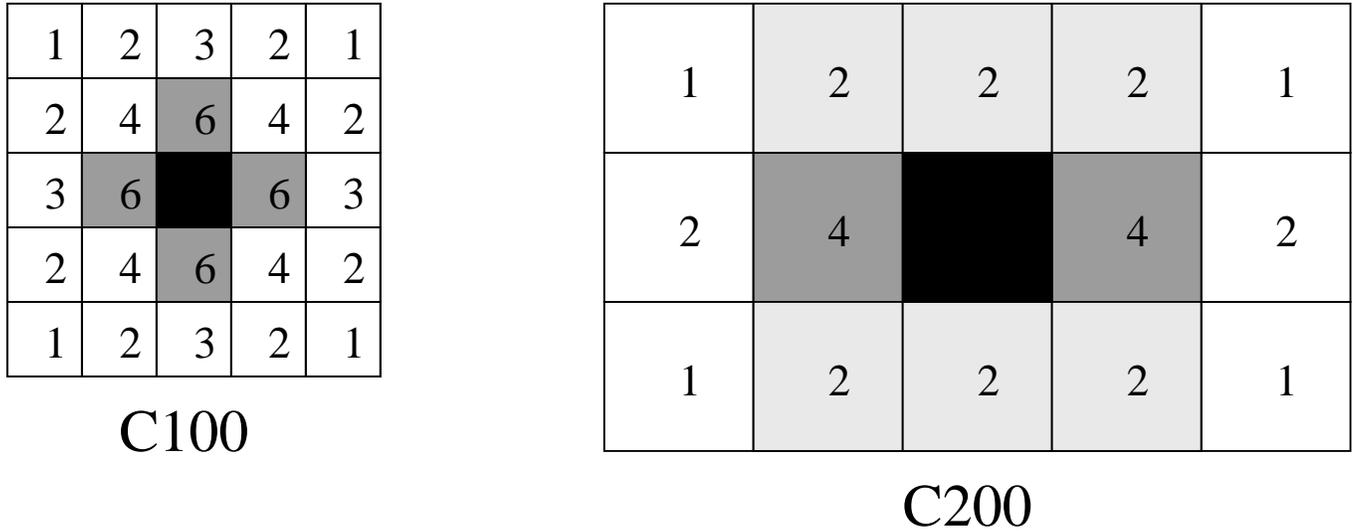}}
  \hfill
  \parbox[b]{18cm}{
  \caption{The layout of the rastermaps on the sky. On the left
  is the C100 layout with a pixel size of $44\arcsec$, on the right is
  the C200 layout with a pixel size of $92\arcsec$. One square
  represents one pixel on the sky and the number equals the number of
  observations of this position. The black square is the source
  position, observed 9 times with C100 and 4 times with C200. The dark
  gray squares are used for background determination, for C200 the
  light gray area is used as average background.}
	}
  \label{rastermaps}
\end{figure*}

At SRD level only deglitching is performed to correct the
remaining cosmic hits or changes in response due to cosmic rays.  A
sigma of 2.5 is used for the object measurements, and 2.0 for the
calibration measurements.  The data are processed to Signal per Chopper
Plateau (SCP) level by averaging all ramp signals over time per raster
point.  At SCP level the data are corrected for reset interval and dark
current is subtracted.  A completely new feature in PIA 8.0 is the
correction for a changing response during the measurements, called
signal linearization.  This is also applied at this level and is a
significant improvement in the detection of faint objects, such as ours. 
After this, only the flux calibration is performed and this introduces
the large 30\% error in the final flux densities.  However, the
detection of a source is independent of this calibration and should be
assessed separately before the calibration is performed.  Thus we use
the SCP data to determine the actual S/N for our objects.  The noise is
calculated from the mean of the errors that PIA gives for the signal on
each raster point.  The calculation of the signal is difficult, since no
pixel average can be used, as the pixels are not corrected for
illumination (flatfielded, in optical terms).  For each pixel a S/N
value is calculated individually and then compared to the median S/N of
all pixels.  In most cases the deviations are averaged out and the
median agrees well with the average S/N for the individual pixels. 
After the calibration the data are at Astronomical APplication level
(AAP) and ready for mapping and background subtraction. 

The flux calibration measurements of ISOPHOT are done inside the
satellite, using two Fine Calibration Sources (FCS1 and FCS2).  For the
mini-maps there are two FCS observations of FCS1, one before and one
after the source observation.  Thus an interpolation is possible to
determine the real responsivity at the time of the source measurement. 
The heating power of the FCS is adjusted to give a signal comparable in
strength to the signal expected from the object flux given in the
initial observing programme.  The observed FCS flux is converted to a
real flux using calibration tables constructed with sources for which
the flux is well determined.  Extrapolation is only possible within a
safe range, the so called soft extrapolation limits.  Less secure
extrapolation is performed up to the hard limits, but outside of these the
default responsivities should be used.  In our case, all measurements
were inside the soft limits and we used the average responsivities from
both FCS measurements for all objects. 

\subsection{ISOPHOT flux determination}

Determining the flux density of the sources requires a good method for
background subtraction. In our raster maps only the central position is
optimally sampled, while the background positions always have less data
points available. Fig.~1 illustrates how many data points are available
for each sky position and which positions are used for the final source
flux density determination. 

\subsubsection{C100 flux densities}

For positions with an oversampling rate of six or more, we calculate the
weighted mean of all the pixels that have observed the position.  Thus
noisy pixels and data taken during the switch-on drift of the detector
are excluded almost automatically.  Putting all the data points in a
spreadsheet provides the possibility to check for other anomalies.  In
most datasets all pixels are well behaved, but sometimes pixel 5 of the
C100 array shows high flux values.  This pixel was excluded from the
datasets where it was found to bias the final flux density
significantly.  For the source observations the first pixel to see the
source (pixel~7) is excluded, since at this position the detector is
severely influenced by the switch-on drift. 

Using only the strongly oversampled positions leaves us with four
background positions and one source position.  In all cases we assume a
flat background and match the four observed background flux densities to
have no large deviations among them.  This involves mainly the exclusion
of one or two bad pixels that are either 'hot' or 'drift' pixels.  In
some cases the data are good enough to use all pixels.  This method thus
provides a possibility to determine a flux without bad pixels and
without losing redundancy. 

The noise is calculated from the PIA errors given on the signals used in
the final flux determination.  This means that when bad pixels are
excluded, the noise goes down as well.  The final value is corrected for
PSF. 

\subsubsection{C200 flux densities}

%
%
\begin{table*}[!ht]
\begin{center}
\leavevmode
\footnotesize
\begin{tabular}[h]{lrrr|lrrr}
\hline \\[-8pt]
Quasar	& 160\,$\mu$m & 90\,$\mu$m & 60\,$\mu$m &
galaxy & 160\,$\mu$m & 90\,$\mu$m & 60\,$\mu$m \\
\hline \\[-8pt]
3C334	& 7	& 13	&  6	&
3C19	& -2	&  2	&  0	\\
3C351	& 14	& 40	& 23	&
3C42	& 2	&  3	&  1	\\
3C323.1 & 4	&  4	&  5	&
3C460	& 17	&  4	&  0	\\
3C277.1 & 4	&  9	&  4	&
3C67	& 4	&  5	&  4	\\
\hline \\
\end{tabular}
\end{center}
\caption{Signal to noise ratios as determined before the calibration
has been applied. See Sect.~3.1 for further details.}
\end{table*}
\normalsize

The same method is applied as for the C100 data, but with one main
difference.  Since the redundancy on these data is much lower, only two
background values can be calculated.  This leaves no possibility to
match them, assuming a flat background.  Therefore, a third value is
calculated using the six positions on either side of the source (see
Fig.~1).  This value is a good estimate of the background: it excludes
any gradients over the source as all positions are distributed
symmetrically around the source.  The only problem is that the upper
three positions are observed by different pixels compared to the lower
three, and the flatfield for the C200 array is not very good. 
Fortunately, most of these errors are averaged out because of the
symmetry and the use of weighted means. 

\subsubsection{PSF corrections}

The PSF of both arrays is larger than one pixel.  This implies that some
of the source flux is in the pixels surrounding the one that observes
the source.  In the final map this means that the positions around the
source see a slightly higher background than the outer positions.  Since
the PSF is well known for point sources, the correction for this is
easy.  In PIA 8.0 there is even a flatfield algorithm in the mapping
procedure that can correct for this (1st quartile flatfielding) and
subsequently the PSF-corrected flux density can be calculated.  One
should take care in using this however, since this method can create
false detections.  Therefore, we only use the mapping as a first
reference and not to calculate actual flux densities.  It is also useful
to compare the background fluxes we obtained with the PIA values; they
match within 10\% for all observations. 

\subsubsection{Upper limits}

The detection of a source is limited by the background flux and the
noise on the signal.  When a source is placed on a high background, even
with low noise on the data it cannot be detected.  The same is true for
a source on a low background with very noisy data.  Thus the calculation
of the upper limits should take both the noise on the data and the
sensitivity of the detector into account.  First the minimum sensitivity
of the detectors is calculated for each wavelength separately.  This is
readily done by dividing the detected fluxes by the background of the
same measurement and taking the smallest value.  Minimum sensitivities
are 3\% for C100 and 1\% for the C200.  This means that on the C100
array a source with only 3/100 of the background flux is still
detectable.  The efficiency is now defined as the product of the minimum
sensitivity and the observed background flux.  For each non-detection
the efficiency was calculated.  This is a first estimate of the flux
that can be detected.  Then we calculate the noise on the signal, using
the same method as described for flux determination.  Subsequently, we
add 2$\sigma$ noise to the efficiency to obtain an upper limit. 
Effectively, this corresponds to $\geq 4\sigma$ upper limits for all
non-detections. 

\subsubsection{Calibration error}

Due to the large calibration errors for faint sources, the fluxes
presented here are to be taken with errors of about 30\% (Lemke et al.,
1996).  These systematical errors are large with respect to the
statistical ones and therefore not included in the errors in Fig.~2 and
Table~3.  The C200 data are taken immediately after the C100 data and
thus any calibration errors due to instrumental effects work in the same
direction.  This implies that the spectral indices are much better
determined than the 30\% calibration error suggests, and their actual
error should be well below 10\%. 

\subsection{VLA and SCUBA data reduction}

The VLA data are reduced using standard AIPS routines for mapping and
self-calibration.  Calibration of the raw interferometer data uses
standard VLA flux calibrators and nearby phase calibrators.  We do not
display the resulting radio maps, as the lack of short spacings prevents
reliable imaging of the extended radio lobes.  Radio source core flux
density values are determined using the MAXFIT procedure. 
\object{3C\,351} caused some problems during the reduction of its 6\,cm
and 2\,cm observations, due to the strong hot spots in this source. 
Using the CLEAN procedure with more fields, this problem was solved.  In
the 6\,cm image of the compact \object{3C\,277.1} the core is convolved
with the lobe on the western side.  Deconvolution with MAXFIT produces a
peak flux of around 40~mJy, but since the lobe is not well approximated
by a gaussian this value is to be taken as an upper limit.  Instead we
use the 6\,cm literature value (Akujor et al. 1991) of 28~mJy.  The
quality of the cm data is very high; the errors listed in Table~3
correspond to the 3$\sigma$ noise in the final radio maps.  For all
sources the errors in the flux density determinations are smaller than
the plot symbols in Fig.~2. 

The reduction of the 7\,mm data appeared to be problematic.  The phases
change so rapidly in time, that in most cases no solution can be found. 
The resulting maps then look like noise maps, although a signal can
clearly be present when plotting the amplitude $uv$-data in a
flux-baseline plot.  Only for \object{3C\,277.1} a phase solution is
found and a flux density can be determined.  In general, for sources
below 100~mJy, the 7\,mm data appeared impossible to reduce.  Thus, the final
flux density values may, in theory, range between 0 and 100~mJy.  In the
case of \object{3C\,334} there is clearly structure in the $uv$
flux-baseline plot, but no phase solution is found.  Since it is not
possible to determine an upper limit in case of phase fitting errors, we
do not include the 7\,mm data in the final analysis, except for
\object{3C\,277.1}. 

%
%
\begin{table*}[!ht]
\label{restab}
\begin{center}
\leavevmode
\footnotesize
\begin{tabular}[h]{lrrrrrrrrr}
\hline \\[-8pt]
Object & $F_{6cm}$ & $F_{2cm}$ & $F_{12mm}$ & $F_{7mm}$ & $F_{2mm}$ &
$F_{850\mu}$ & $F_{160\mu}$ & $F_{90\mu}$ & $F_{60\mu}$ \\
\hline \\[-8pt]
\object{3C\,334} & 167.7 $\pm$ 0.8 & 112.3 $\pm$ 0.4 & 91.4 $\pm$ 0.7 & -- &
 20 $\pm$ 10 & 15 $\pm$ 8 & 55 $\pm$ 20 & 59 $\pm$ 10 & 86 $\pm$ 22 \\
\object{3C\,19} & -- & -- & -- & -- & -- & -- & $\leq$65 & $\leq$25 & $\leq$50 \\
\object{3C\,351} & 9.3 $\pm$ 0.3 & 5 $\pm$ 1 & $\leq$5.0 & -- & -- & -- &
 110 $\pm$ 15 & 145 $\pm$ 10 & 211 $\pm$ 23 \\
\object{3C\,42} & -- & -- & -- & -- & -- & -- & $\leq$45 & $\leq$45 & $\leq$ 80\\
\object{3C\,323.1} & 38 $\pm$ 2 & 43 $\pm$ 1 & 45 $\pm$ 1 & -- & $\leq$14 &
 $\leq$12 & 30 $\pm$ 19 & 33 $\pm$ 9 & 49 $\pm$ 18 \\
\object{3C\,460} & -- & -- & -- & -- & -- & -- & 96 $\pm$ 25 & 14 $\pm$ 7 &
 $\leq$51 \\
\object{3C\,277.1} & 28* & 22.7 $\pm$ 0.7 & 21.5 $\pm$ 1.0 & 9.6 $\pm$ 1.8&
 $\leq$17 & -- & 56 $\pm$ 14 & 30 $\pm$ 10 & 50 $\pm$ 25 \\
\object{3C\,67} & -- & -- & -- & -- & -- & -- & 73 $\pm$ 50 & 44 $\pm$ 8 &
 46 $\pm$ 18 \\
\hline \\
\end{tabular}
\end{center}
\caption{Flux densities and errors in mJy as plotted in Fig.~2. The
  value marked with * is from Akujor et al. (1991).}
\end{table*}
\normalsize

The SCUBA data are reduced using the SCUBA Data Reduction Facility
(SURF) and Kernel Application Package (KAPPA).  This is a relatively
straightforward reduction, since SCUBA is a bolometer array, allowing
immediate comparison of the middle bolometer with the surrounding ones. 
Not all sources were observed at all bands; undetected sources have
upper limits, and only \object{3C\,334} has been marginally detected
($\sim2.5\sigma$) in both bands.  The 450\,$\mu$m data do not
show any source signal, due to the large atmospheric extinction at this
wavelength.  The upper limits are too high to put any useful constraints
on our data and are therefore omitted from Table~3 and Fig.~2.  The
detections at 850\,$\mu$m and 2\,mm are marginal at best, but the upper
limits provide strong constraints on the non-thermal contamination of
the far-infrared emission.

\section{Results}

%
%
\begin{figure*}
  \resizebox{18cm}{!}{\includegraphics{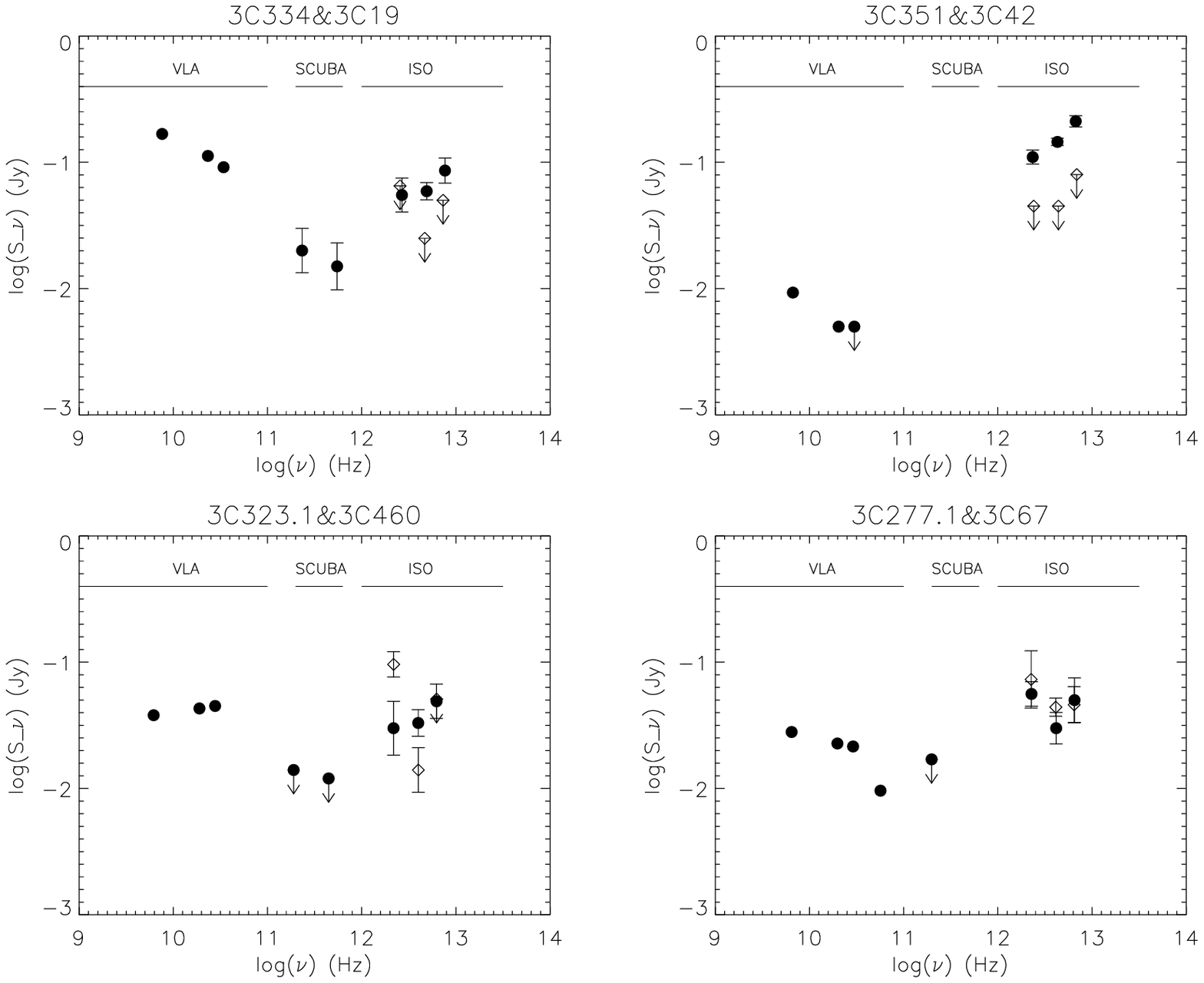}}
  \hfill
  \parbox[b]{18cm}{	
    \caption{The complete spectra -- in their emitted frame --
   for the four pairs observed with ISO. Filled circles are quasar data,
 open diamonds are radio galaxy data. The wavelength range of the
 different instruments is shown in the plots by the horizontal bars.}
    \label{figaph}}
\end{figure*}	

Since the calibration from detector signal to flux density introduces a
substantial error, we make a clear distinction between the detection of
a source and the determination of the observed flux.  In all cases the
detected source has a well determined flux and the objects where the
flux measurent is difficult have low S/N.  Thus the method is in
principle consistent.  In practice, however, it is possible that an
object is detected but has no flux determination because the calibration
errors are too large.  In Table~2 we present the S/N values before
calibration.  We have assumed a detection when the S/N exceeds 3. 

In Table~3 we present the resulting flux densities, only with the final
1$\sigma$ intrinsic noise.  This value is in general nearly equal to the
predicted 30\% calibration error.  In Fig.~2 we have again plotted only
the intrinsic noise, providing better insight into the value of the
detection and the spectral indices. 

It is clear from Table~3 and Fig.~2 that two pairs exhibit clear flux
density differences.  In the pair 3C\,334 and 3C\,19, the radio galaxy
is not detected in any band, whereas the quasar has three
solid detections (lowest S/N is 6).  Since the upper limits take the
difference of the backgrounds into account, the radio galaxy is
definitely fainter than its paired quasar.  In the pair consisting of
quasar 3C\,351 and radio galaxy 3C\,42, the former is unusually bright
in the far-infrared (as already observed by IRAS), whereas the radio
galaxy has no detections.  As the radio core in 3C\,351 is relatively
weak, the contribution of any beamed non-thermal far-infrared component
can be ruled out.  The pair consisting of quasar 3C\,323.1 and radio
galaxy 3C\,460 displays a difference at the two short wavelengths.  The
radio galaxy is not detected at 60\,$\mu$m and has a flux density well
below that of the quasar at 90\,$\mu$m.  However, an interesting change
occurs at long wavelengths; the 160\,$\mu$m flux density of 3C\,460
exceeds the 3C\,323.1 value.  Below we will argue that cool dust
associated with star formation is likely responsible for this luminous
160\,$\mu$m radiation. 

In contrast, the pair consisting of quasar 3C\,277.1 and radio galaxy
3C\,67 has entirely comparable far-infrared output.  Remarkably, this is
the pair of Compact Steep Spectrum sources (CSS), objects believed to be
young FRIIs where the radio emission is still confined within the host
galaxy.  Both objects are well detected, with S/N values around 5.  Also
remarkable in this pair is the high 160\,$\mu$m flux for both the radio
galaxy and the quasar; the 160\,$\mu$m flux densities exceed the
60\,$\mu$m values.  The only other sample object where this is observed
is the radio galaxy 3C\,460.  It should be noted however, that 3C\,67
shows a broad component in H$\alpha$, and hence should be classified as
a broad-line radio galaxy.  We will return to this issue below. 

In all cases the 160\,$\mu$m point seems to be higher than what is
expected from extrapolating the 90 and 60\,$\mu$m data with optically
thin grey body models.  This could be evidence for cold dust in the host
galaxies.  Unless contrived colour corrections within the ISOPHOT bands
are large and negative, this seems to be a real effect.  Because we do
not know the intrinsic shape of the spectral energy distribution, we
choose not to apply any colour corrections.  In general the far-infrared
spectra of the quasars indicate a positive spectral index at higher
frequencies (i.e., shorter infrared wavelengths) and a flattening
towards the lower frequencies{\footnote{We adopt $F_{\nu} \propto \nu
^{\alpha}$}}.  Unfortunately we have no data at mid-infrared wavelengths
permitting to determine spectral turnovers in that range, as have been
observed in nearby Seyfert galaxies and radio-quiet quasars (Polletta
\& Courvoisier 1998, Andreani 1998, Klaas et al. 1998).  ISO archive data
show however that the short wavelength flux densities for 3C\,351 are
well below our 60\,$\mu$m point, implying a peaked spectrum, with the
peak somewhere between 60 and 25\,$\mu$m. 

\subsection{Far-infrared luminosities}

In order to better compare the observations, we have calculated the
luminosities in the ISOPHOT bands: Table~4 lists the resulting values,
in Watts\footnote{We adopt $H_0=75$ km\,s$^{-1}$\,Mpc$^{-1}$, $q_0=0.5$
and $k\Lambda =0$ throughout this paper}.  For the spectral index in the
K-correction, we used the actual values as observed with ISOPHOT, except
for $L_{160}$, where we used $\alpha=0$, assuming that we observe the
peak of the cold dust emission with $T \sim 20$\,K as observed in normal
galaxies (Bianchi et al. 1999) and Seyferts (Rodr{\'\i}guez Espinosa et
al. 1996).  To make a valid comparison between the radio galaxies and
the quasars in the present sample, we calculated the median luminosities
at all wavelengths for the different classes.  It appears that, with the
exception of the CSS pair, the sample quasars are brighter than the
radio galaxies at all wavelengths. The median luminosities for the
latter always include at least two upper limits, which implies that the
actual values for this class are even lower.  Thus we can solidly
conclude that the quasars are brighter by at least a factor of 1.5 at
60\,$\mu$m, which strengthens results from IRAS samples and the Haas et
al.  (1998) measurements.  However, even at longer wavelenghts we find
that the median luminosity of the quasars is higher than the median
luminosity of the radio galaxies, the latter of which being an upper
limit.  At 90\,$\mu$m the quasars are a factor of 1.5 brighter than the
radio galaxies, at 160\,$\mu$m this is a factor of 1.3, excluding
3C\,460, which is clearly an exceptional object.  In summary, the
supergalactic size quasars and radio galaxies of our admittedly small
ISOPHOT sample differ in their far-infrared output, in the sense that
the quasars are brighter than the radio galaxies at all wavelengths by a
factor $\sim1.5$. 

\subsection{Nature of the infrared emission}

%
%
\begin{table*}[!ht]
\begin{center}
\leavevmode
\footnotesize
\begin{tabular}[h]{lccc|lccc}
\hline \\[-8pt]
Quasar & log($L_{160}$) & log($L_{90}$) & log($L_{60}$) &
Galaxy & log($L_{160}$) & log($L_{90}$) & log($L_{60}$) \\
\hline \\[-8pt]
3C334	& 37.68	& 38.00	& 38.11	&
3C19	& $\leq$37.64 & $\leq$37.65 & $\leq$37.75 \\
3C351	& 37.64	& 38.00	& 38.16	&
3C42	& $\leq$37.33	& $\leq$37.72	& $\leq$37.86	\\
3C323.1 & 36.82	& 37.22	& 37.35	&
3C460	& 37.34 & 37.12	& $\leq$37.37 \\
3C277.1 & 37.25	& 37.44	& 37.47	&
3C67	& 37.34	& 37.49	& 37.51	\\
\hline \\[-8pt]
Median QSR & 37.45 & 37.72 & 37.79	&
Median RG  & $\leq$37.34 & $\leq$37.57 & $\leq$37.63	\\
\hline \\
\end{tabular}
\end{center}
\caption{Luminosities ($log(\nu L_{\nu})$)calculated from the ISO
observations, units are in W. The spectral indices are calculated from
the observations, while
for non-detections we have used the extreme value, to obtain an upper
limit for the luminosity. For the L$_{160}$ we used $\alpha$=0. We
adopt $F_{\nu} \propto \nu^{\alpha}$.}
\end{table*}
\normalsize

From Fig.~2 it is clear that the extrapolation of the radio core
spectrum to the infrared underpredicts the observed infrared flux
densities.  This implies that (relativistically beamed) nonthermal
radiation is not contributing significantly in the present quasars.  The
only exception could be 3C\,334, where a maximum of about 10\% beamed
radiation might be present in the infrared, confirming previous results
(van Bemmel et al. 1998).  Still, this is not enough to explain the
difference between 3C\,19 and 3C\,334, given that their flux densities
appear to differ by about a factor of two.  The contribution of beamed
non-thermal emission is well below 2\% in the other cases.  This implies
that the bulk of the observed far-infrared emission in the sample
objects must be of thermal origin. 

\subsection{Dust mass estimates}

Having established the thermal origin of the infrared emission, we can
estimate the dust mass responsible for the radiation.  Because the high
160\,$\mu$m points render a single component grey body fit impossible,
we fitted the observations with a two temperature grey body.  Since this
requires a four component fit to three points, we kept the temperatures
constant throughout the sample to obtain a uniform estimate of the dust
masses.  The temperatures are chosen to be 75\,K for the warm component,
as we are not sensitive to much higher temperatures due to the limited
wavelength coverage, and 20\,K for the cold component (as observed from
cold cirrus in nearby galaxies, Bianchi et al. 1999, Rodr{\'\i}guez
Espinosa et al. 1996).  Only for 3C\,460 did the fit require a slightly
lower temperature for the cold component of 18\,K.  Calculating the dust
masses with fixed temperatures gives an average mass of
$3\times10^5$~$M_{\odot}$ for the warm component and
$1.5\times10^8$~$M_{\odot}$ for the cold one.

\section{Discussion}

With the absolute flux uncertainties of ISOPHOT and the small sample
under study, we are not able to constrain the model predictions from
either the unification theory or the dust models for Seyfert galaxies. 
Any statistical analysis will be severly biased by the individual
characteristics of the objects in the sample. We therefore choose to
only compare the median luminosities of the two classes. Below we
discuss shortcomings of the present samples as well as various
alternative interpretations of the ISOPHOT data.

\subsection{Compact versus extended sources}

The pair containing subgalactic size CSS sources stands out; they have
(within the errors) identical infrared output.  While this could --
most interestingly -- be a general property of dust emission from the
host galaxies of young radio sources, it could also relate to the fact
that we have matched a broad-line radio galaxy to a quasar.  3C\,67 has
been reported to have broad H$\alpha$ (Eracleous \& Halpern 1994), but
is otherwise in its spectral characteristics clearly different from
strong quasars such as 3C\,277.1.  It will be interesting to examine the
ISO archive in order to see if the compact sources in general have
comparable infrared output, whether classified as quasar or radio
galaxy, or if we have made a mismatch in this case. 

\subsection{Size and AGN luminosity differences}

Inspection of Table~1 shows that although the observed paired objects
match in redshift and radio luminosity, they differ in projected linear
radio size, the quasars being signifantly larger (with the exception of
the CSS pair).  Within the framework of the unification model, the size
difference would be even more pronounced, given the larger deprojection
factor for the quasars.  This fact, although purely coincidental, might
have severe consequences for the interpretation of our results.  On the
basis of model evolutionary tracks in the P--D diagrams (e.g.  Kaiser et
al. 1997), larger objects with the same radio power should have more
powerful jets.  Within this scenario, the AGN strengths, and consequently
the far-infrared output of our sample quasars, could be at least a factor
of two larger as compared to the radio galaxies, which matches our
observations. 

In addition, within the so-called receding torus model and given the
possibility of some scatter in the optical-UV AGN luminosity, the
average quasar may be a factor $\sim$2 more luminous than the average
radio galaxy, when drawn from a sample having one and the same radio
luminosity (Simpson 1998). 

Along these lines, we may have paired powerful quasars with somewhat
less powerful radio galaxies.  This explanation could in principle be
tested by intercomparison of the luminosities in the [OIII] and/or [OII]
emission lines, arising in the circumnuclear narrow-line region.  While
not enough [OII]$\lambda3727$ data are available in the literature, the
[OIII]$\lambda\lambda$4959,5007 line luminosity will be inconclusive, as
the latter is known to be anisotropic (Hes et al. 1996), due to
dust extinction within the narrow line region (di Serego Alighieri et
al. 1997, Baker 1997). 

Analysis of the FIR properties of much larger samples, covering larger
parameter space will be needed to address these effects.

\subsection{Host galaxy contribution}

Strong evidence for a host galaxy contribution comes from 3C\,460, where
a cold dust mass of 2.4$\times10^8$~M$_{\odot}$ is observed.  This
amount of dust is most likely due to star formation in the host galaxy,
which could be related to a close companion galaxy at 1.5$\arcsec$
separation (de Koff et al. 1996).  A distinction between torus,
starburst and cold cirrus dust was already made by Rowan-Robinson
(1995).  From 60$\mu$m data on PG quasars there is indeed evidence that
this emission is at least in part produced by star formation (Clements,
2000). Thus dust in the host galaxy can dilute the emission from a
(warmer) dust torus. 

\subsection{Variability of the non-thermal component}

Non-thermal flares which can occur within days have been observed in
blazars, and should be observable in all AGN with a jet-axis close to
the line of sight.  These flares peak in the millimeter range (Brown et
al. 1989), and can be responsible for strong variability down to
far-infrared wavelengths.  We do not cover the spectral range where
these flares are observed, nor do we have the time resolution to detect
any variability.  Thus variability cannot be excluded by our data. 

\subsection{Anisotropy in the torus emission}

All models for emission from a dust torus heated by a central AGN are
based on parameters deduced from Seyfert galaxies.  A straightforward
application to the stronger radio galaxies and quasars might not be
possible, as these sources have generally much stronger central cores
and different hosts.  Their tori might be much thicker and denser and
thus emit optically thick at the observed wavelengths. 

\section{Conclusions}

Observations of pairs of radio-loud quasars and radio galaxies confirm
the previously reported far-infrared excess in quasars and indicate that
this excess extends up to restframe wavelengths of $\sim130\,\mu$m. 
However, the origin of this excess remains unclear and may lie in the
biases introduced in our limited sample.  It has been shown that more 
than 98\% of the observed far-infrared emission is of thermal origin,
with the exception of 3C\,334, where non-thermal emission might
contribute up to 10\%.  The 160\,$\mu$m data points are always higher
than the expected flux densities for grey body emission, which suggests
the presence of a cold component in all objects.  In 3C\,460 we report
the discovery of a comparatively large cold dust content, which could be
related to interaction with a close companion. 

In order to explain the observed difference in infrared output, we
discuss several possibilities, such as size difference, host
contributions, non-thermal flares, anisotropy of the torus, and a true
AGN power difference.  With the present quality and limited amount of
data we are not able to discriminate between either of these.  We stress
the need for better modelling of torus emission for 3C-like objects,
which is at present not available for the wavelengths under
consideration.  Further analysis of ISOPHOT data on similar objects will
be undertaken to enlarge the sample and better constrain the models. 

\begin{acknowledgements}

We would like to thank all ISOPHOT people who helped during the hard
reduction process, especially Carlos Gabriel, Bernhard Schulz and Ren\'e
Laureijs, for important advice on the data reduction. Thanks to Martin
Haas, who provided a lot of tips and tricks. We further acknowledge
conversations with Chris Carilli, Eric Hooper, Maria Poletta, Belinda
Wilkes, Thierry Courvoisier and Rolf Chini. Also thanks to Ronald Hes
for his initial involvement in this project, and to Jane
Dennett-Thorpe, Mark Neeser
and Xander Tielens for useful comments and discussions. We acknowledge
expert and careful reading by the referee and the editor.

PIA is a joint development by the ESA Astrophysics division and the
ISOPHOT consortium. The National Radio Astronomy Observatory is a
facility of the National Science Foundation operated under cooperative
agreement by Associated Universities, Inc. The James Clerk Maxwell
Telescope is operated by The Joint Astronomy Centre on behalf of the
Particle Physics and Astronomy Research Council of the United Kingdom,
the Netherlands Organisation for Scientific Research, and the National
Research Council of Canada. 

\end{acknowledgements}

\newpage


\begin{thebibliography}{}

\bibitem[1991]{Ak}
Akujor C.E., Spencer R.E., Zhang F.J., et al., 1991, MNRAS 250, 215

\bibitem[1998]{An}
Andreani P., 1998, in {\it 'The universe as seen by ISO}', 20 -- 23
October 1998, Paris, ESA SP-427, p.857

\bibitem[1993]{ant93}
Antonucci R., 1993, 
ARA\&A 31, 473

\bibitem[1997]{bak97}
Baker J. C., 1997, MNRAS 286, 23

\bibitem[1989]{pdb89}
Barthel P.D., 1989,
ApJ 336, 606

\bibitem[1999]{Bi}
Bianchi S., Alton P.B., Davies J.I., 1999, in {\it 'ISO beyond point
sources'}, 14 -- 17 September, 1999, VILSPA 

\bibitem[1994]{bog94}
Bogers W.J., Hes R., Barthel P.D., Zensus J.A., 1994,
AAS 105, 91

\bibitem[1994]{brid94}
Bridle A.H., Hough D.H., Lonsdale C.J., et al., 1994,
AJ 108, 766

\bibitem[1989]{brwn89}
Brown L.M.J., Robson E.I., Gear W.K., et al., 1989,
ApJ 340, 129

\bibitem[1999]{Cl}
Clements D.L., 2000,
MNRAS 311, 833

\bibitem[1997]{koff97}
de Koff S., Baum S.A., Sparks W.B., et al., 1996, 
ApJS 107, 621

\bibitem[1998]{wimdev98}
de Vries W.H., O'Dea C.P., Baum S.A., et al., 1998,
ApJ 503, 156

\bibitem[1997]{serego97}
di Serego Alighieri S., Cimatti A., Fosbury R.A.E., Hes R.,
1997, A\&A 328, 510

\bibitem[1994]{}
Eracleous M., Halpern J.P., 1994
ApJS 90, 1

\bibitem[1974]{fr74}
Fanaroff B.L., Riley J.M., 1974,
MNRAS 167, 31P

\bibitem[1990]{fant90}
Fanti R., Fanti C., Schilizzi R.T., et al., 1990, A\&A 231, 333

\bibitem[1997]{fern97}
Fernini I., Burns J.O., Perley R.A., 1997,
AJ 114, 2292

\bibitem[1994]{gran94}
Granato G.L., Danese L., 1994,
MNRAS 268, 235

\bibitem[1998]{haas98}
Haas M., Chini R., Meisenheimer K., et al., 1998,
ApJ 503, L109

\bibitem[1994]{heck94}
Heckman T.M., O'Dea C.P., Baum S.A., Laurikainen E., 1994,
ApJ 428, 65

\bibitem[1995]{hes95}
Hes R., Barthel P.D., Hoekstra H., 1995,
A\&A 303, 8

\bibitem[1996]{hes96}     
Hes R., Barthel P.D., Fosbury R.A.E., 1996,
A\&A 313, 423  

\bibitem[1997]{hoek97}
Hoekstra H., Barthel P.D., Hes R., 1997,
A\&A 319, 757

\bibitem[1988]{imp88}
Impey C.D., Neugebauer G., 1988, AJ 95, 307

\bibitem[1977]{jenk77}
Jenkins C.R., Pooley G.G., Riley J.M., 1977,
MemRAS 84, 61

\bibitem[1997]{Ka}
Kaiser C.R., Dennett-Thorpe J., Alexander P., 1997,
MNRAS 292, 723

\bibitem[1998]{Kl}
Klaas U., Haas M., Schulz B., 1998, in {\it 'The universe as seen
by ISO}', 20 -- 23 October 1998, Paris, ESA SP-427, p.901

\bibitem[1980]{laing80}
Laing R.A., 1980,  
MNRAS 193, 439

\bibitem[1996]{lem96}
Lemke D., Klaas U., Abolins J., et al., 1996,
A\&A 315, L64

\bibitem[1997]{ogle97}
Ogle P.M., Cohen M.H., Miller J.S., et al., 1997,
ApJ 482, L37

\bibitem[1992]{pier92}
Pier E.A., Krolik J.H., 1992,
ApJ 401, 99

\bibitem[1998]{Po}
Polletta M., Courvoisier T.J.-L., 1998, in {\it 'The universe as
seen by ISO}', 20 -- 23 October 1998, Paris, ESA SP-427, p.953

\bibitem[1991]{rawl91}
Rawlings S., Saunders R., 1991, 
Nat. 349, 138

\bibitem[1996]{rodrig96}
Rodr\'{\i}guez Espinosa J.M., P\'erez Garc\'{\i}a A.M., Lemke D.,
Meisenheimer K., 1996, A\&A 315, L129

\bibitem[1995]{rowanr95}
Rowan-Robinson M., 1995,
MNRAS 272, 737

\bibitem[1995]{sang95}
Sanghera H.S., Saikia D.J., L\"udke E., et al., 1995, 
A\&A 295, 629

\bibitem[1998]{simps98}
Simpson C., 1998,
MNRAS 297, L39

\bibitem[1991]{stock91}
Stockton A., Ridgway S.E., 1991,
AJ 102, 488

\bibitem[1999]{tadh99}
Tadhunter C.N., Packham C., Axon D.J., et al., 1999,
ApJ 512, L91

\bibitem[1994]{ueno94}
Ueno S., Koyama K., Nishida M., Yamauchi S., Ward M., 1994, 
ApJ 431, L1

\bibitem[1995]{urry95}
Urry C.M., Padovani P., 1995,
PASP 107, 803

\bibitem[1998]{ivb98}
van Bemmel I.M., Barthel P.D., Yun M.S., 1998,
A\&A 334, 799

\bibitem[1999]{Wil}
Willott C.J., Rawlings S., Blundell K.M., Lacy M., 1999
MNRAS 309, 1017

\bibitem[1997]{wink97}
Wink J.E., Guilloteau S., Wilson T.L., 1997,
A\&A 322, 427

\end{thebibliography}
\end{document}